\documentclass[aps, prb, superscriptaddress, twocolumn, 10pt, amsfonts, amsmath, amssymb,
 reprint,
]{revtex4-2}

\usepackage{graphicx}
\usepackage{dcolumn}
\usepackage{bm}
\usepackage{hyperref}

\usepackage{lipsum}
\usepackage{braket}
\usepackage{xcolor}
\usepackage{amsmath}
\hypersetup{
    colorlinks,
    linkcolor={blue!80!black},
    citecolor={blue!80!black},
    urlcolor={blue!80!black}
}
\usepackage[normalem]{ulem}

\def\ee{\mathrm{e}}
\newcommand{\Up}{{\uparrow}}
\newcommand{\Dn}{{\downarrow}}

\begin{document}


\title{Majorana modes in quantum dots coupled via a floating superconducting island}


\author{Rubén Seoane Souto}
\thanks{These authors contributed equally to this work.}
\affiliation{Instituto de Ciencia de Materiales de Madrid (ICMM),
Consejo Superior de Investigaciones Cient\'ificas (CSIC),
Sor Juana In\'es de la Cruz 3, 28049 Madrid, Spain}

\author{Virgil V. Baran}
\thanks{These authors contributed equally to this work.}
\affiliation{Faculty of Physics, University of Bucharest, 405 Atomi\c stilor, RO-077125, Bucharest-M\u agurele, Romania}
\affiliation{``Horia Hulubei" National Institute of Physics and Nuclear Engineering, 30 Reactorului, RO-077125, Bucharest-M\u agurele, Romania}
\affiliation{Center for Quantum Devices, Niels Bohr Institute, Copenhagen University, 2100 Copenhagen, Denmark}

\author{Maximilian Nitsch}
\affiliation{Division of Solid State Physics and NanoLund, Lund University, S-22100 Lund, Sweden}

\author{Lorenzo Maffi}
\affiliation{Dipartimento di Fisica e Astronomia “G. Galilei”,
Universit\`a degli Studi di Padova, I-35131 Padova, Italy}
\affiliation{Istituto Nazionale di Fisica Nucleare (INFN), Sezione di Padova, I-35131 Padova, Italy}

\author{Jens Paaske}
\affiliation{Center for Quantum Devices, Niels Bohr Institute, Copenhagen University, 2100 Copenhagen, Denmark}

\author{Martin Leijnse}
\affiliation{Division of Solid State Physics and NanoLund, Lund University, S-22100 Lund, Sweden}

\author{Michele Burrello}
\affiliation{Center for Quantum Devices, Niels Bohr Institute, Copenhagen University, 2100 Copenhagen, Denmark}
\affiliation{Niels Bohr International Academy, Niels Bohr Institute, Copenhagen University, 2100 Copenhagen, Denmark}
\affiliation{Dipartimento di Fisica dell’Università di Pisa and INFN, Largo Pontecorvo 3, I-56127 Pisa, Italy}

\date{\today}

\begin{abstract}
Majorana modes can be engineered in arrays where quantum dots (QDs) are coupled via grounded superconductors, effectively realizing an artificial Kitaev chain. Minimal Kitaev chains, composed by two QDs, can host fully-localized Majorana modes at discrete points in parameter space, known as Majorana sweet spots. Here, we extend previous works by theoretically investigating a setup with two QDs coupled via a floating superconducting island. We study the effects of the charging energy of the island and the properties of the resulting minimal Kitaev chain. We initially employ a minimal perturbative model, valid in the weak QD-island coupling regime, to derive analytic expressions for the Majorana sweet spots and the splitting of the ground state degeneracy as a function of tunable physical parameters. The conclusions from this perturbative approximation are then benchmarked using a microscopic model that explicitly describes the internal degrees of freedom of the island. Our work shows the existence of Majorana sweet spots, even when the island is not tuned at a charge-degeneracy point. In contrast to the Kitaev chains in grounded superconductors, these sweet spots involve a degeneracy between states with a well-defined number of particles.
\end{abstract}

\maketitle

\section{Introduction}

Electrostatic interactions constitute a key aspect in the engineering of a large variety of quantum devices, including quantum dots (QDs), single-electron transistors and Cooper-pair boxes. The effects of charging energy have played a crucial role also in the design of many hybrid semiconductor-superconductor systems (see, for instance, the recent reviews \cite{Prada_review,flensberg2021engineered,Souto_arXiv2024}) and provide important tools for the study of subgap states in nanodevices fabricated with the goal of realizing topological superconductors  
\cite{Higginbotham_NatPhys2015,Albrecht_Nature2016,van_Heck_PRB2016,Albrecht_PRL2017,Shen_NatComm2018,Farrell_PRL2018,Hansen_PRB2018,Sabonis_APL2019,Whiticar_NatComm2020,Vaitiekenas_Science2020,Vaitiekenas_PRB2022,Wang_SciAdv2022,Valentini_Nature2022,Razmadze_PRB2024,Valentini_arXiv2024}.

Among the hybrid systems, artificial Kitaev chains, composed by QDs coupled via superconductors, have emerged as a promising platform to realize Majorana modes~\cite{Leijnse_PRB2012,Sau_NatComm2012,Fulga_NJP2013,Liu_PRL2022}, eventually reaching the topological regime from a bottom-up approach~\cite{Svensson_arXiv24,Luethi_arXiv24}.
In their minimal form, these chains are composed by two QDs coupled via a short grounded superconducting segment that mediates crossed Andreev reflection (CAR) and elastic cotunneling (COT)~\cite{Leijnse_PRB2012,Sau_NatComm2012}. These minimal Kitaev chains feature the so-called poor man's Majorana modes (PMMs) at isolated sweet spots in parameter space where CAR and COT amplitudes are equal.

Experiments in QDs coupled via grounded superconductors have demonstrated the onset of zero-energy states with transport and spectral properties compatible with the PMM interpretation in two~\cite{Dvir2023,Zatelli_arXiv2023,Haaf2024} and three QD Kitaev chains~\cite{bordin_arXiv2024,Haaf_arXiv2024}. The main strategy to realize PMMs in these experiments relies on Andreev subgap states in the central superconductor, that dominate the coupling between the QDs. By controlling the energy of these states, it is possible to tune the relative amplitude of CAR and COT, which, in turn, allows for reaching the Majorana sweet spot~\cite{Liu_PRL2022,Tsintzis2022,Bordin_PRX2023,Bordin_PRL2024}.
The high level of tunability reached makes artificial Kitaev chains a promising alternative to demonstrate the properties of Majorana modes, like Majorana fusion and braiding~\cite{Liu_PRB2023,tsintzis2023roadmap,Boross_PRB2024}, and their applications, including qubits based on fermion parity~\cite{tsintzis2023roadmap,Pino_PRB2023,Pan_arXiv2024}.

In this work, we analyze the low-energy properties of two QDs coupled via a floating superconductor. We examine the role of the charging energy of this superconducting island and show that the system features sweet spots with Majorana modes well localized in the QDs, both at and away from the charge-degeneracy point. We first present a perturbative model based on the effective CAR and COT coupling between spin-polarized QDs, Sec. \ref{sec:minimalModel}. Within this minimal model, the effect of the island charging energy is similar to that of an inter-QD charging energy considered in Refs.~\cite{Samuelson_PRB24,Brunetti_PRB2013}. The predictions of the perturbative model are benchmarked in Sec. \ref{Sec:microscopic} by using a microscopic model that includes spin and Coulomb interactions of the QDs and explicitly describes the central island, based on the recently developed surrogate model~\cite{Baran_PRB2023,Baran2024Jun}.
In the appendices we derive analytic expressions for the energy renormalization of the QDs and amplitudes of CAR and COT processes, and we further analyse the effects of charging energy and multiple subgap states in the floating superconductor.

Our analysis is complementary to the recent investigations of electrostatic interactions in the QDs, which have shown their beneficial effect to improve the quality of PMMs \cite{Tsintzis2022,Samuelson_PRB24,Brunetti_PRB2013} and their role to determine triple degeneracies in the energy spectra of these systems \cite{Bozkurt_arXiv2024}. 

The effects of the charging energy have also been analyzed in superconducting nanowires aimed at more traditional realizations of Majorana modes. The interplay of electrostatic interactions with superconductivity in these systems has been at the core of many theoretical proposals concerning Majorana-mediated teleportation \cite{Fu_PRL2010}, topological qubits~\cite{Hassler_NJP2011,Plugge_NJP2017}, the initialization of topological systems~\cite{Aasen_PRX2016,Nitsch_PRB2022}, Majorana fusion~\cite{Souto_SciPost2022,Nitsch_PRB2024}, and braiding~\cite{van_Heck_NJP2012,Hyart_PRB2013,Wu_arXiv2024}. This work points to the emergence of PMMs in interacting systems that display the potential of reproducing these devices and phenomena in tunable hybrid QD--superconductor platforms. Such setups will open the door for the realization of qubits based on PMMs and the observation of unique Majorana properties, including Majorana teleportation and the topological Kondo effect \cite{Nitsch24}.

\begin{figure}
\includegraphics[width=0.45\textwidth]{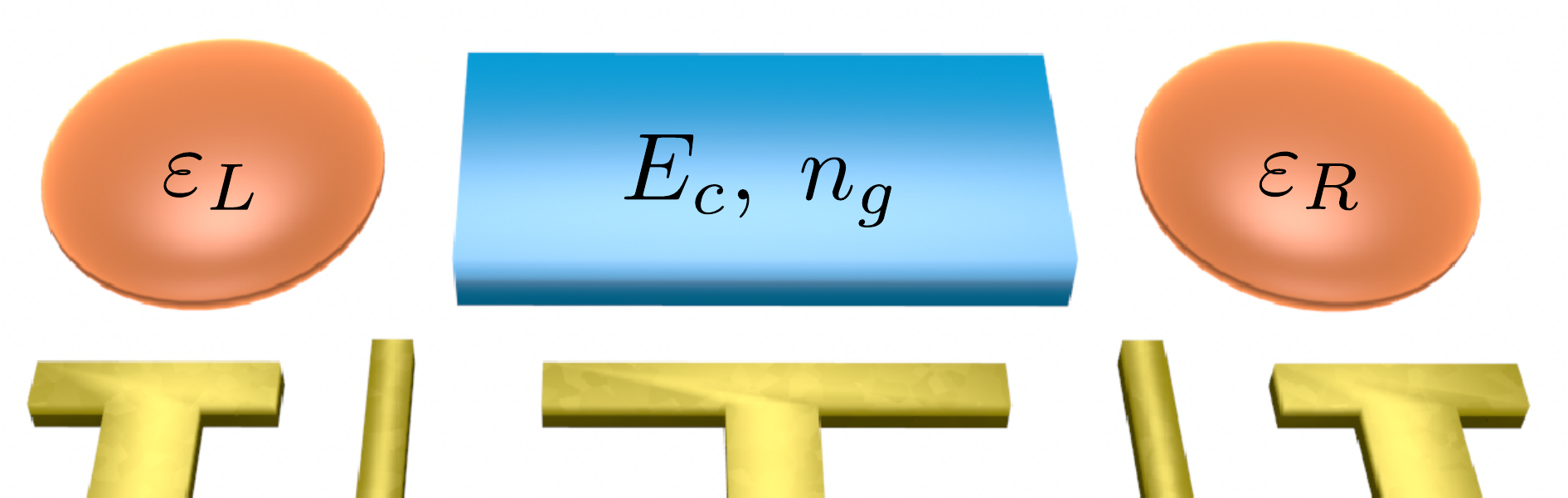}
    \caption{Sketch of the setup considered, where two QDs (orange) couple via a floating superconducting island (blue). Both the QD energy levels ($\varepsilon_\nu$) and the charge offset $n_g$ on the superconductor can be controlled with gate electrodes (gold).}
\label{fig_1}
\end{figure}

\section{Minimal model} \label{sec:minimalModel}

The devices we consider are composed by two tunable QDs coupled through tunneling to a central superconducting island (see Fig.~\ref{fig_1}). These setups can be realized in both nanowires \cite{Dvir2023,Bordin_PRL2024,Zatelli_arXiv2023} and two-dimensional electron gases \cite{Haaf2024} proximitized by a superconductor.

To obtain analytic results, we initially consider a simplified model where the central superconductor only mediates the non-local coupling between the QDs~\cite{Leijnse_PRB2012}
 \begin{equation}
 	H= H_{\rm d} + H_T + H_I\,.
  \label{Eq:Ham1}
 \end{equation}

To pinpoint the basic features of the system, we consider spin-polarized QDs that can effectively be described by
\begin{equation} \label{HQDs}
	H_{\rm d}= \sum_{\nu}\epsilon_\nu d_{\nu}^\dagger d_\nu\,,
\end{equation} 
where $\nu=L,R$ indicates the QD with creation (annihilation) operator $d_{\nu}^\dagger$ ($d_{\nu}$); we label by $\epsilon_\nu$ their effective energy levels. In Appendix \ref{sec:pert} we derive an expression for $\epsilon_\nu$ from a microscopic model: $\epsilon_\nu$ displays in general a subleading dependence on the charge in the superconducting island. For the sake of simplicity, here we neglect this correction, whose effects are summarized in Appendix \ref{sec:pert}.

We consider that the central superconductor mediates CAR and COT processes between the QDs, with respective amplitudes $t_{\rm COT}$ and $\Delta_{\rm CAR}$
\begin{equation} \label{Heff}
    H_{T}= t_{\rm COT}\,d^{\dagger}_L d_R+\Delta_{\rm CAR}\,d^{\dagger}_Ld^{\dagger}_R \ee^{-i{\hat{\phi}}}+{\rm H.c.}\,,
\end{equation}
where we have integrated the continuum of quasiparticles in the superconductor. Here, we use the canonically conjugate Cooper pair number and phase operators $\hat{N}_c$ and $\hat{\phi}$, $[\hat{N}_c,\ee^{i\hat{\phi}}]=\ee^{i\hat{\phi}}$, where $\ee^{\pm i\hat{\phi}}$ adds/removes one Cooper pair from the superconducting condensate.

We describe the electrostatic interactions of the superconducting island as 
\begin{equation} \label{H_I1}
    H_{I}= E_c(2\hat{N}_c-n_g)^2\,,
\end{equation}
where we disregarded excitations in the island, considering the energy of the quasiparticles of the proximitized central region $\tilde{\Delta}\gg E_c,\,t_{\rm COT}\,,\Delta_{\rm CAR}$. Therefore, all electrons in the island form Cooper pairs. Here, $\hat{N}_c$ is the number operator of Cooper pairs in the island, and $n_g$ a charge offset that can be controlled using electrostatic gates.

Under these conditions, the quantum state of the system can be characterized by three numbers, labelling the occupation of each of the two QDs and the superconducting island. We adopt the notation  $\ket{n_L\,n_R,N_c}$ with $n_\nu$ denoting the charge occupation of QD $\nu$. Two of these families of states display an even fermionic parity and have the form $\ket{0\,0,N_c}$ and $\ket{1\,1,N_c-1}$ for a total number of particles $2N_c$. The other two have odd fermionic parity and they correspond to $\ket{1\,0,N_c}$ and $\ket{0\,1,N_c}$ for a total number of particles $2N_c+1$. 

In this basis and with the above assumptions and approximations, the Hamiltonian in Eq.~\eqref{Eq:Ham1} in the even parity sector becomes
\begin{equation} \label{He}
 	H_e=\left[
	\begin{matrix}
	E_c(2N_c-n_g)^2 & \Delta_{\rm CAR} \\
	\Delta_{\rm CAR} & E_c(2N_c-2-n_g)^2+\epsilon_L+\epsilon_R
	\end{matrix}\right]\,.
\end{equation}
Under the condition
\begin{equation} \label{constr1}
    \epsilon_L=\epsilon_R= 2E_c\left(2N_c-1-n_g \right) \equiv -\delta E_c/2\,,
\end{equation}
the eigenstates of the even parity Hamiltonian acquire the following form
\begin{equation}
    \ket{e^{\pm}}=\frac{1}{\sqrt{2}}\left(\ket{0\,0,N_c} \pm \ket{1\,1,N_c-1}\right). \label{estate}
\end{equation}
The states $\ket{e^{\pm}}$ display an equal superposition of both even configurations of the QD occupations, condition that follows from considering localized Majorana modes in the QDs.
Under the constraint \eqref{constr1}, the energies of the two eigenstates with even fermionic parity result:
\begin{equation}
E^{\pm}_{\rm e}(2N_c) = E_c(2N_c-n_g)^2 \pm\Delta_{\rm CAR}\,.
\end{equation}
In order to engineer zero-energy PMMs, these energies must be equal to the eigenenergies of the odd sector, whose Hamiltonian, for $2N_c+1$ electrons, becomes
\begin{equation} \label{Ho}
 	H_o=\left[
	\begin{matrix}
	\epsilon_L & t_{\rm COT} \\
	t_{\rm COT} & \epsilon_R
	\end{matrix}\right]+E_c(2N_c-n_g)^2\,.
\end{equation}
Under the condition $\epsilon_L=\epsilon_R$, the odd eigenstates have the form
\begin{equation}
\ket{o^{\pm}}=\frac{1}{\sqrt{2}}\left(\ket{1\,0,N_c} \pm \ket{0\,1,N_c}\right)\,, \label{ostate}
\end{equation}
that is an equal superposition of each QD being empty and occupied, as required for the Majorana sweet spot.
Under the constraint \eqref{constr1}, their energies are
\begin{equation}
E^{\pm}_{\rm o}(2N_c+1) = E_c(2N_c-n_g)^2 \pm t_{\rm COT}-\delta E_c/2\,,
\end{equation}
Therefore, in order to fulfill the degeneracy of the ground states in the even and odd sectors with $2N_c$ and $2N_c+1$ electrons, we must impose the condition 
\begin{equation}
    \left|\Delta_{\rm CAR}\right|=\left|t_{\rm COT}\right|+\delta E_c/2\,.
    \label{Eq:delta_condition}
\end{equation}
This expression predicts the existence of Majorana sweet spots, not only at the charge-degeneracy points, $n_g=(2k+1)$, but also inside the Coulomb-blockaded valleys. Equations \eqref{constr1} and \eqref{Eq:delta_condition} determine the sweet spots that appear in this interacting system when, on one side, the energy of the QDs are shifted to compensate for the electrostatic effects in the even parity sector, and, on the other, the relative amplitudes for CAR and COT are modified to ensure the Majorana properties of the quantum states. We note that a similar condition was derived in Ref.~\cite{Samuelson_PRB24} for two QDs coupled via non-local charging energy, which leads to analogous physics when considered within the minimal model.

We now focus on the properties of the system as it is detuned from the Majorana sweet spot. The energy splitting between the even and odd ground states is given by 
\begin{eqnarray}\label{Eq:dE_splitting}
    \delta E_{e-o}=\frac{1}{2}\left(\delta E_c+\sqrt{4 \left|t_{\rm COT}\right|^2+(\epsilon_L-\epsilon_R)^2}\right.\nonumber\\
    \left.-\sqrt{4\left|\Delta_{\rm CAR}\right|^2+(\delta E_c +\epsilon_L+\epsilon_R)^2}\right)\,.
\end{eqnarray}
We first focus on the situation where one QD is detuned while the second is at resonance, {\it i.e.} $\epsilon_R=-\delta E_c/2$ and $\epsilon_L=-\delta E_c/2+\delta \epsilon_L$. In this case we obtain
\begin{eqnarray} \label{Eq:dE_splitting2b}
    \delta E_{e-o}=\frac{1}{2}\left(\delta E_c+\sqrt{4 \left|t_{\rm COT}\right|^2+\delta \epsilon_L^2}\right.\nonumber\\
    \left.-\sqrt{4\left(\left|t_{\rm COT}\right|+\delta E_c/2\right)^2+\delta\epsilon_L^2}\right)\,,
\end{eqnarray}
where we enforced the expression~\eqref{Eq:delta_condition} for $\Delta_{\rm CAR}$ at the sweet spot.
For $|\delta \epsilon_L|\ll |t_{\rm COT}|$, we find
\begin{equation} \label{Eq:dE_splitting3}
    \delta E_{e-o}\approx\frac{\delta\epsilon_L^2}{8}\left(\frac{1}{\left|t_{\rm COT}\right|}-\frac{1}{\left|\Delta_{\rm CAR}\right|}\right)\,.
\end{equation}
At the charge degeneracy points ($n_g=2N_c-1$ and thus $\delta E_c=0$), the ground state remains doubly degenerate when detuning the energy of one of the QDs, similar to what happens in the non-interacting situation~\cite{Leijnse_PRB2012}. In this situation, the PMM in the detuned QD spills into the other QD. It means that the ground state wavefunction still has Majorana character in the detuned QD, but with a reduced weight. The situation is different away from the charge degeneracy points, where $|t_{\rm COT}|\neq|\Delta_{\rm CAR}|$ at the sweet spot. In this case, detuning one of the QDs results in a quadratic splitting of the ground states energy. This occurs because each PMM has a finite weight on both QDs when the energy of one of the QDs is detuned, with $n_g$ shifted away from the charge degeneracy point.

Changes of $t_{\rm COT}$ or $\Delta_{\rm CAR}$ result in a linear ground state splitting, similar to the original PMM result with a grounded superconductor~\cite{Leijnse_PRB2012}. In the same way, deviations of $n_g$ from the sweet spot would lead to a linear splitting of the ground states in Eq. \eqref{Eq:dE_splitting} corresponding to $\delta E_{e-o} \approx 2E_c \delta n_g$, with $\delta n_g$ being the shift of the charge offset.

In the previous analysis, we focus on the charging energy of the central island and we 
disregarded further electrostatic interactions. In the setup shown in Fig.~\ref{fig_1}, the cross capacitance between the QDs and the superconducting island can be significant, affecting the properties of the system. Therefore, it  must be compensated for a precise determination of the Majorana sweet spots. This is in not an issue in conventional Kitaev chains where the grounded superconductors feature a vanishing charging energy.

The cross capacitance betweent the QDs and the island shifts the energy of the system according to
\begin{equation}
    H_{\rm xc}=2\sum_\nu U_{ I\nu} n_{\nu}{\hat{N}_c}\,,
\end{equation}
where $n_\nu=d^\dag_\nu d_\nu$ represents the charge on the QD $\nu$ and $U_{I\nu}$ is the non-local Coulomb interaction arising from the cross capacitance. Here, we neglected the offset charge $n_g$ as it corresponds to a trivial shift of the QD energies, $\epsilon_\nu$. 

To determine the corrections to the tuning conditions \eqref{constr1} and \eqref{Eq:delta_condition}, it is convenient to introduce the dressed QD energy $\tilde\epsilon_\nu=\epsilon_\nu+2U_{I\nu}N_c$. In this way, the odd Hamiltonian can be written as Eq.~\eqref{Ho}, using the renormalized levels, $\tilde \epsilon_\nu$. On the other hand, the even state $\ket{11,N_c-1}$ displays a different charge of the central island. Therefore, the Hamiltonian~\eqref{He} of the even sectors must be modified not only by using the renormalized QD levels, $\tilde{\epsilon}_\nu$, but also including the further shift $-2 (U_{IL}+U_{IR})$ in its second diagonal entry. Consequently, the conditions for a Majorana sweet spot now read as 
\begin{equation}\label{condition1_XC}
\tilde\epsilon_L=\tilde\epsilon_R=-\delta E_c/2+U_{IL}+U_{IR}\,,
\end{equation}
and
\begin{equation}\label{condition2_XC}
    \left|\Delta_{\rm CAR}\right|=\left|t_{\rm COT}\right|+\delta E_c/2-U_{IL}-U_{IR}\,.
\end{equation}
Therefore, a sweet spot appears under a different tuning of parameters, needed to compensate for the energy shift $H_{\rm xc}$. In this analysis we have ignored the cross capacitance effect between the outer QDs, as we expect them to be smaller than the ones between the QDs and the island.  These terms would lead to a shift of $\delta E_c$ in Eqs. \eqref{condition1_XC} and \eqref{condition2_XC}, as discussed in Ref.~\cite{Samuelson_PRB24} for two QDs coupled to a grounded superconductor.
For the sake of simplicity, in the following we will neglect all cross capacitance interactions.

\section{Microscopic calculations}\label{Sec:microscopic}
We now numerically test the results of the previous section in a more microscopic model that considers explicitly the electron spin and the fermionic degrees of freedom in the central island connecting the QDs, and which includes Coulomb interactions on the QDs. The full Hamiltonian preserves the form of Eq.~\eqref{Eq:Ham1}, with the QD Hamiltonian being
\begin{equation}
\label{ham_dots}
    H_{d}=\sum_{\nu,\sigma}\varepsilon_{\nu\sigma} d_{\nu\sigma}^\dagger d_{\nu\sigma}+U_\nu n_{\nu\uparrow}n_{\nu\downarrow},
\end{equation}
where $\sigma=\uparrow,\downarrow$ denotes the electron spin and $\varepsilon_{\nu\uparrow}=\varepsilon_{\nu\downarrow}-E_Z\equiv\varepsilon_\nu$, with $E_Z$ being the Zeeman splitting in the QDs. Here, $U_\nu$ indicates the local charging energy in the QDs, which plays no role in the spinless minimal model of Sec.~\ref{sec:minimalModel}.

\begin{figure}
\includegraphics[width=\columnwidth]{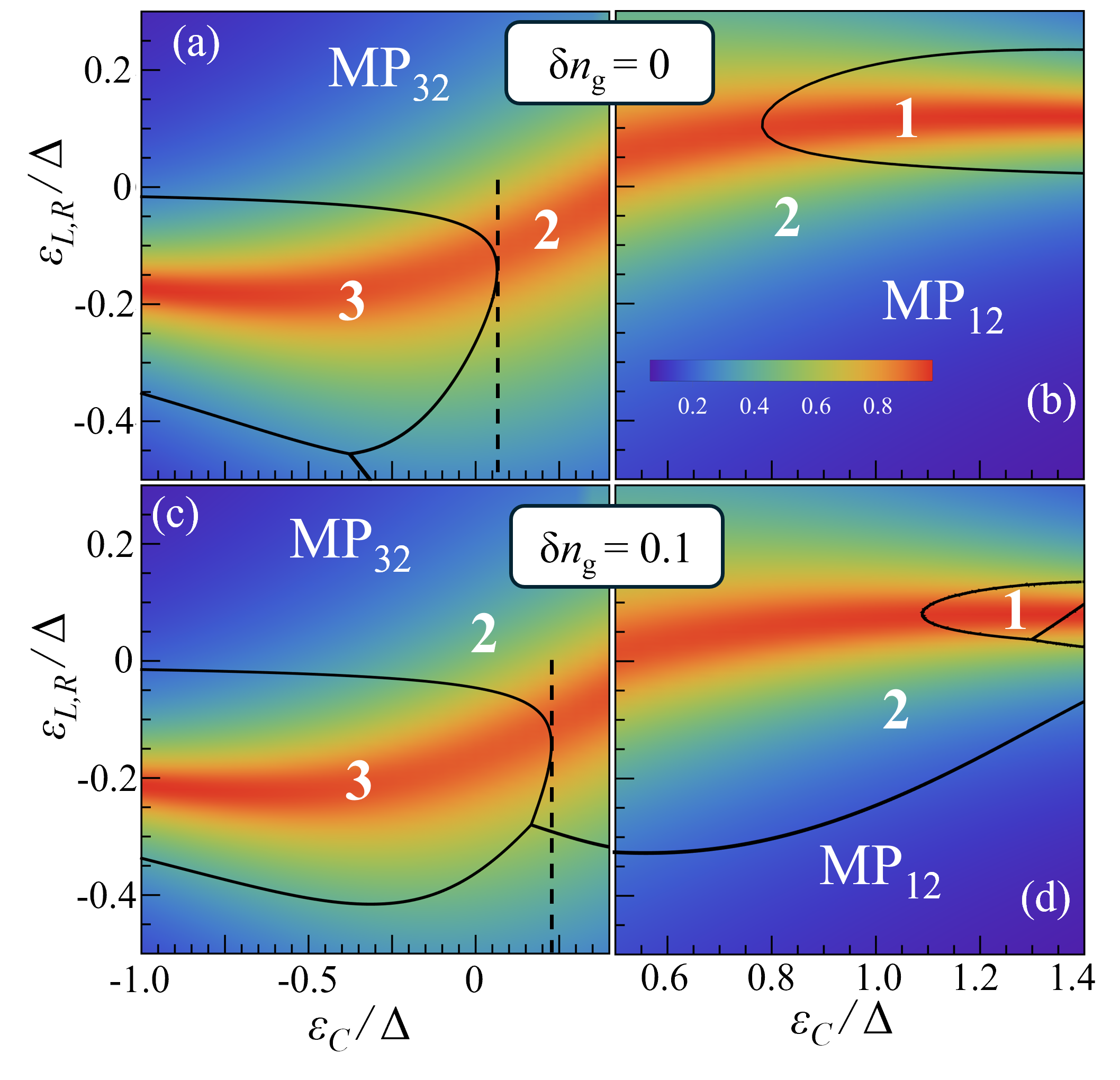}
    \caption{Majorana polarization at $\delta n_g=n_g-n_{g,0}=0$ (top row) and $\delta n_g = 0.1$ (bottom row), with $n_{g,0}=1$ denoting an odd charge reference. The black lines mark the charge-degeneracy conditions, defining regions where the ground state has a well-defined number of electrons. Here MP$_{ij}$ denotes the Majorana polarization computed between states with total number of electrons $i$ and $j$. The white labels indicate the ground state charge in each sector. The system parameters are $U_{\text{L,R}}=5\Delta$, $t_{L,R}=0.5\Delta$, $t^{\rm SO}_{L,R}=0.1\Delta$, $E_{Z;{L,R}}=1.5\Delta$, $E_c=0.2\Delta$. The dashed vertical lines indicate the cuts analyzed in Fig. \ref{fig_3}. 
    }
\label{fig_2}
\end{figure}

Present-day superconducting hybrid islands, e.g. realized as epitaxial superconductor-semiconductor systems, show a dense continuum of states above the bulk gap, which is induced by the metallic superconducting shell, and just a small number of proximitized semiconductor sub-gap levels to which the outer QDs couple. We model the central superconducting island using the recently developed surrogate model~\cite{Baran_PRB2023,Baran2024Jun}. In the main text, we consider the case where the superconducting island hosts a single sub-gap state. In this limit, the surrogate model is equivalent to the so-called zero-bandwidth model. Additional sub-gap states lead only to quantitative changes in the results, see App.~\ref{app_surrogate}. We thus consider the central island Hamiltonian \cite{Baran2024Jun}
\begin{equation}\label{H_SI}
\begin{aligned}
    {H}_{\text{I}} =& \sum_{\sigma=\uparrow \downarrow} \varepsilon_c c_{\sigma}^{\dagger} c_{\sigma}- (\Delta c_{\uparrow}^{\dagger} c_{\downarrow}^{\dagger} \ee^{-i{\hat{\phi}}}+\text{H.c.})+E_c \left(\hat{N}_{\rm I}-n_g\right)^2,
    \end{aligned}
\end{equation}
where $\hat{N}_{\rm I}=\sum_\sigma c^\dagger_\sigma c_\sigma+2 \hat{N}_c$ is the total charge in the island, with $c_\sigma$ being the annihilation operator for electrons in the island's subgap states, see App.~\ref{app_BCS} for details.

For simplicity, we neglect in Eq. \eqref{H_SI} the Zeeman energy of the electrons in the proximitized region of the semiconductor. We assume the related $g$-factor to be quenched by the coupling to the superconductor and thus much smaller than in the QDs. We observe, however, that the introduction of a further Zeeman splitting in Eq. \eqref{H_SI} can be easily accounted for also in the perturbative approach presented in App. \ref{sec:pert} and does not lead to qualitative differences.

Finally, the tunneling between the island and the QDs is described by
\begin{equation}
\label{ham_hop}
    H_{T}=\sum_{\sigma\nu}\left( t_\nu d^{\dagger}_{\nu\sigma} c_\sigma+{s_\nu} s_\sigma t^{\rm SO}_\nu d^{\dagger}_{\nu\sigma} c_{\bar{\sigma}}+{\rm H.c.}\right),
\end{equation}
where $t$ and $t^{\rm SO}$ are the spin-preserving and spin-flip tunneling amplitudes; the latter corresponds to a rotation of the spin around the $\hat{y}$ axis when an electron tunnels from the dots to the island and is consistent with a Rashba term determined by an out-of-plane gradient of the chemical potential. In Eq. \eqref{ham_hop}, $\bar{\sigma}$ denotes the opposite spin to $\sigma$, $s_\sigma=\pm1$ for up and down spin, and $s_\nu=\pm1$. 

In the limit  $E_Z\to\infty$ and $t, t^{\rm SO} << \Delta - E_c$, the low energy features of this microscopic Hamiltonian become equivalent to the minimal model studied in Sec.~\ref{sec:minimalModel}, where the amplitudes for CAR and COT are given in App.~\ref{sec:pert} as a function of the microscopic parameters in Eqs. \eqref{H_SI} and \eqref{ham_hop}.

\begin{figure}[t!]
\includegraphics[width=\columnwidth]{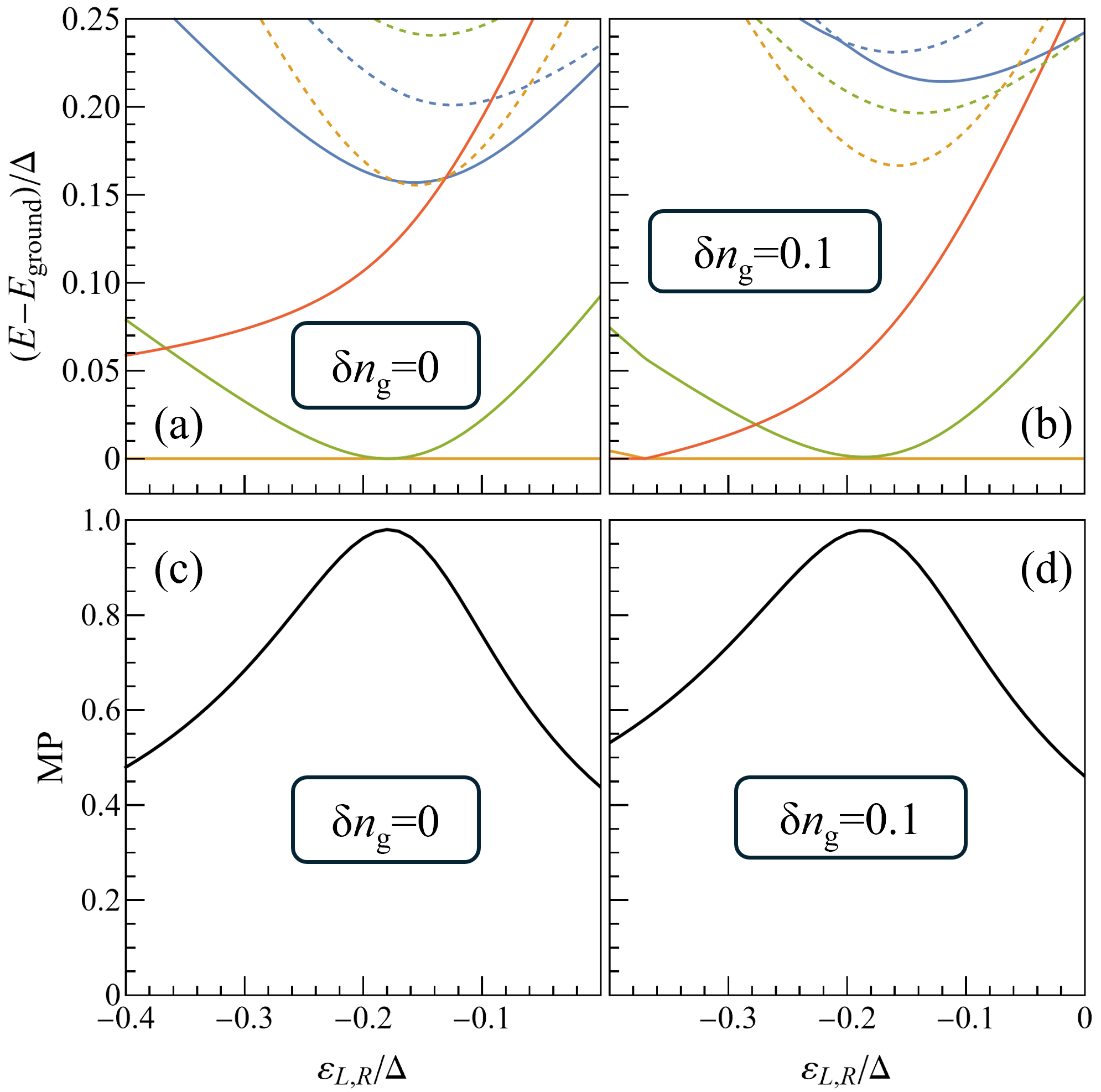}
    \caption{Energy gaps (a,b) and Majorana polarization (c,d)  along the dashed cuts in Figs.~\ref{fig_2}(a) and (c). The colored continuous (dashed) curves correspond to the lowest (first excited) state in each particle-number sector. Panels (a,c) corresponds to a charge degeneracy point $\delta n_g=0$, and panels (b,d) to the detuning $\delta n_g=0.1$.}
\label{fig_3}
\end{figure}

To analyze the sweet spots that appear in the microscopic model, we use the Majorana polarization (MP), defined consistently with the gauge choice in Eq. \eqref{ham_hop} as \cite{Aksenov2020,Tsintzis2022,Samuelson_PRB24}
\begin{equation}\label{MP}
    {\rm MP}(\nu)=\frac{\sum_\sigma (w^2_{\nu\sigma}-z^2_{\nu\sigma})}{\sum_\sigma (w^2_{\nu\sigma}+z^2_{\nu\sigma})}\,.
\end{equation}
In our charge-conserving Hamiltonian, the lowest energy states with even and odd parity differ by 1 electron. We define the MP by adopting $w_{\nu\sigma}=\langle N|(\ee^{i\hat{\phi}}d_{\nu\sigma}+d_{\nu\sigma}^\dagger)|N-1\rangle$ and $z_{\nu\sigma}=\langle N|(\ee^{i\hat{\phi}}d_{\nu\sigma}-d_{\nu\sigma}^\dagger)|N-1\rangle$, where $\ket{N}$ denotes the lowest-energy states with total charge $N$ in the system, such that MP is invariant under a standard gauge transformation $d_{\nu\sigma}\to \ee^{i\alpha}d_{\nu\sigma}$ and $\ee^{i\hat{\phi}}\to \ee^{i\hat{\phi}-i2\alpha}$. Values of MP close to $\pm1$ characterize situations where a local PMM maps the even to the odd electron parity ground state.

In Fig.~\ref{fig_2} we consider the MP in the proximity of a charge degeneracy point, $n_g=n_{g,0}$, where $n_{g,0}$ is an odd integer number. In particular, we depict the polarization ${\rm MP}_{ij}$ calculated between the ground states with total electron numbers $i$ and $j$.
The solid black lines denote the conditions for ground state degeneracy between the even and the odd fermion parity ground states. Majorana sweet spots correspond to discrete points in parameter space where $|{\rm MP}|\approx1$ and the ground state is degenerate. At the charge degeneracy points, $n_g=n_{g,0}$, the system exhibits sweet spots for the same conditions as in the non-interacting situation, see Figs.~\ref{fig_2}(a,b) and Ref.~\cite{Tsintzis2022Nov} for comparison. Here, the difference is that states with different fermion numbers are split due to the electrostatic repulsion in the central superconductor. 

The comparison of upper and lower panels in Fig. \ref{fig_2} displays a general stability of the main features of the ground states landscape and the related MPs under a small detuning of the induced charge $n_g$ away from the charge degeneracy point $n_{g,0}$. As expected, by increasing $n_g$ from $n_{g,0}$ to $n_{g,0}+0.1$, the region in parameter space with ground states displaying a lower particle number, $N=n_{g,0}$, shrinks, whereas, in general, states with $N=n_{g,0}+2$ electrons acquire a lower energy. Under the moderate detuning $\delta n_g=0.1$, the system still features sweet spots characterized by a high MP and energy degeneracy between states with even and odd fermion parity (see Fig. \ref{fig_3}).

Counter intuitively, shifting the charge offset $n_g$ away from the charge-degeneracy point at $n_{g,0}$ generally results in an improvement of MP, see Figs.~\ref{Ap:fig_1b} and \ref{Ap:fig_2}. This is due to an increase of the quasiparticle excitation energy in the island, which leads to a suppression of their average occupation, thus limiting the deviations from the perturbative model introduced in Sec.~\ref{sec:minimalModel}; see Figs.~\ref{Ap:fig_1b} and \ref{Ap:fig_2} in  App.~\ref{App::c}.

\begin{figure}[t!]
\includegraphics[width=\columnwidth]{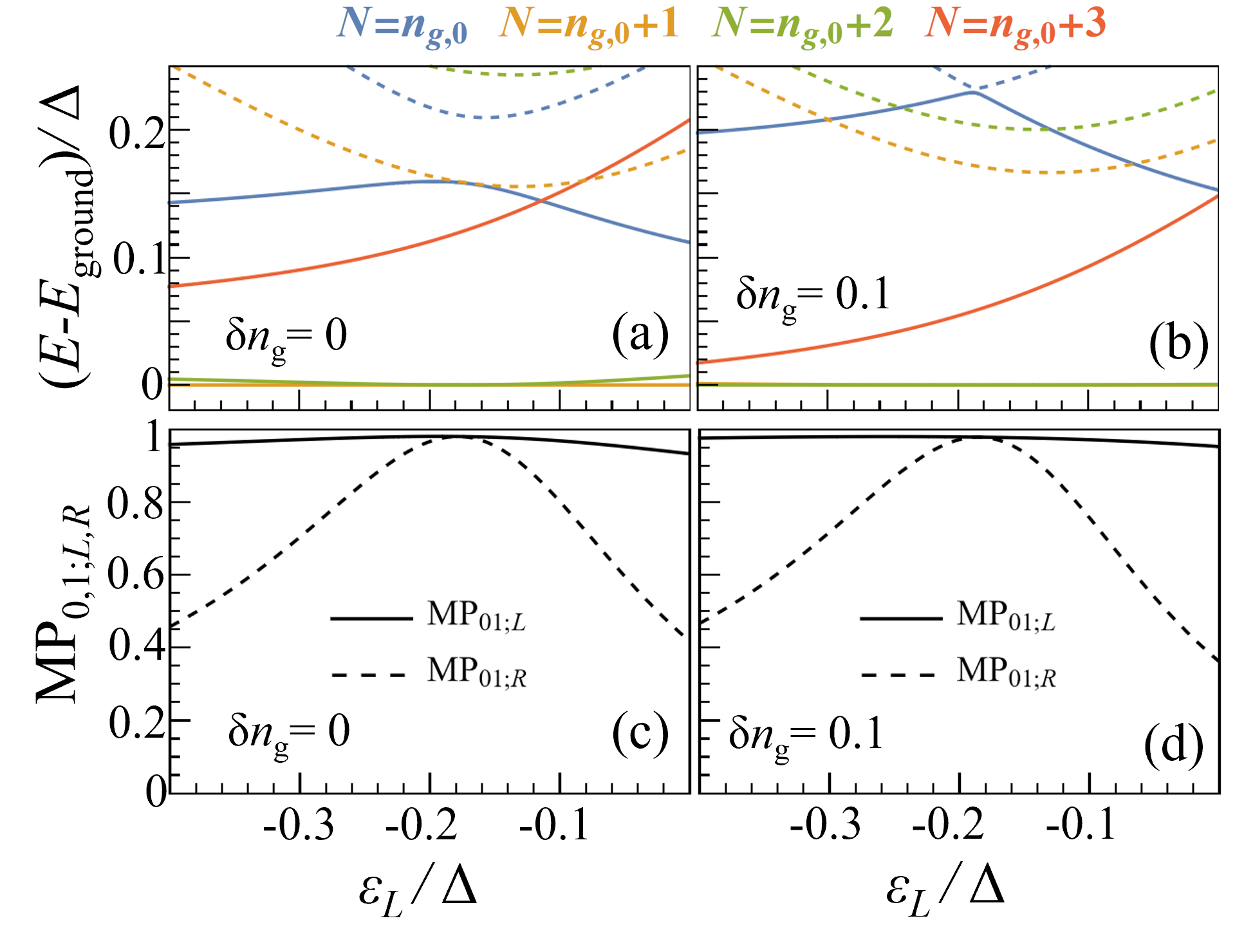}
    \caption{Energy gaps (a,b) and Majorana polarization (c,d) vs the energy level of the left QD. The colored continuous (dashed) curves correspond to the lowest (first excited) state in each particle-number sector. For $\delta n_{g}=0$ the sweet spot is located at $\varepsilon_L=\varepsilon_R=-0.1793\Delta$, $\varepsilon_C=-0.2829\Delta$. For $\delta n_{g}=0.1$ the sweet spot is located at $\varepsilon_L=\varepsilon_R=-0.1865\Delta$, $\varepsilon_C=-0.104\Delta$. The other parameters are $U_{\text{L,R}}=5\Delta$, $t_{L,R}=0.5\Delta$, $t^{\rm SO}_{L,R}=0.1\Delta$, $E_{Z;{L,R}}=1.5\Delta$, $E_c=0.2\Delta$, and $n_{g,0}$ is an odd integer. The splitting of the ground state energy in panel (b) is so small that the green and orange lines overlap.}
\label{fig_4}
\end{figure}

We next study the ground state splitting as a function of the different parameters, see Figs.~\ref{fig_3} and \ref{fig_4}. Just as in the minimal model [Eq. \eqref{Eq:dE_splitting}], the two degenerate ground states with even and odd fermion parity split quadratically when detuning both the QDs, irrespective of the value of $n_g$,  see Fig.~\ref{fig_3} (green curves). This behavior is similar to the one found in minimal Kitaev chains with a grounded superconductor when detuning both QDs~\cite{Leijnse_PRB2012}.

A second kind of detuning concerns the parameters of the central superconducting island. Detuning either $\varepsilon_c$ or the charge offset $n_g$ renormalizes the amplitudes for CAR and COT, leading to a linear splitting of the two ground states from the sweet spot defined by Eq. \eqref{Eq:delta_condition} (not shown), similar to the minimal model in Sec. \ref{sec:minimalModel}.

\begin{figure}[t!]
\includegraphics[width=\columnwidth]{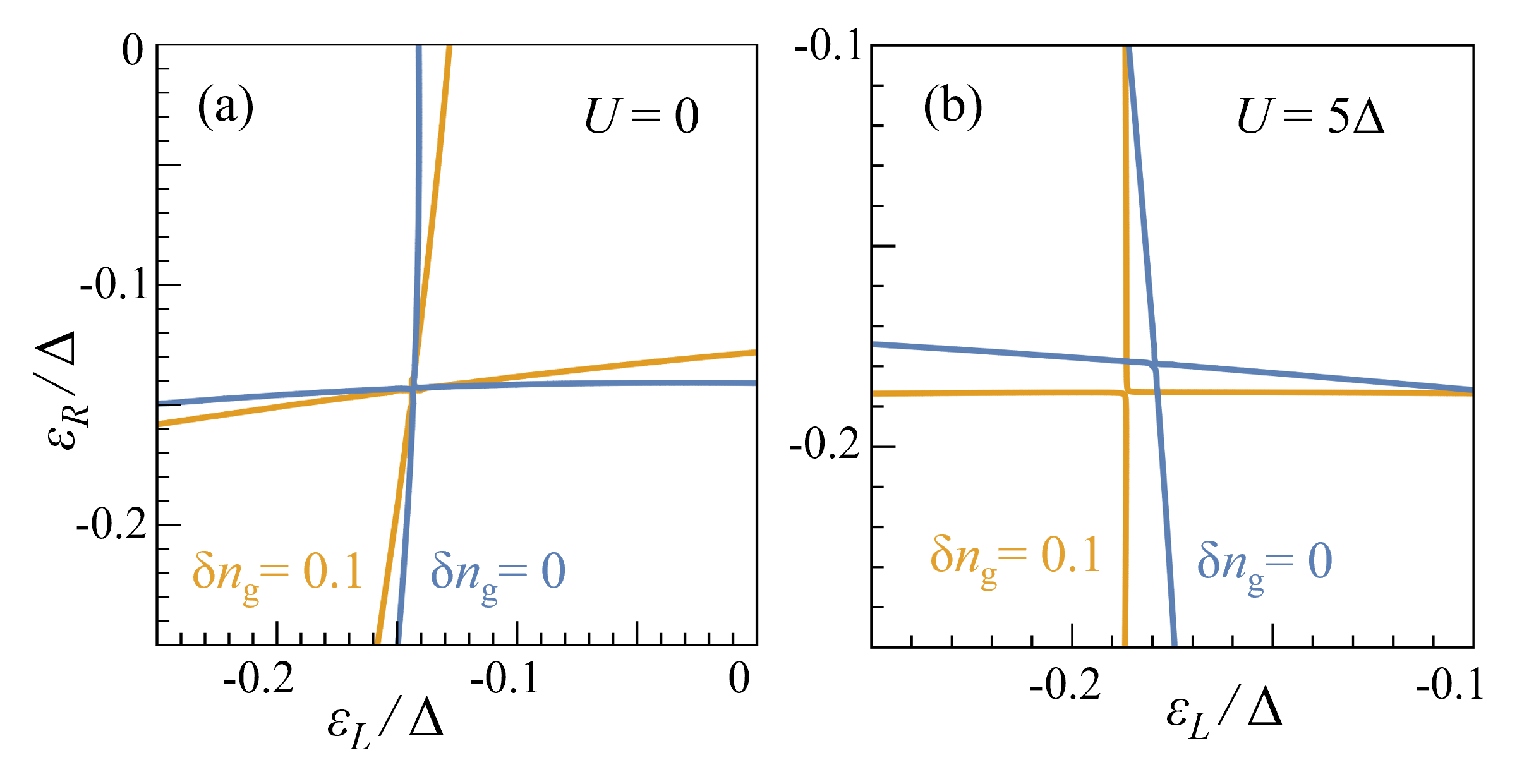}
    \caption{Lines denoting the parameters with ground state degeneracy as a function of the energy of the QDs for $U=0$ (a) and $U=5\Delta$ (b). We show results for $\delta n_g=0$ (blue) and $\delta n_g=0.1$ (orange). The Majorana sweet spots correspond to the intersection point between the two lines with the same color. The other parameters are the same as in Figs. 2,3,4.} 
\label{fig_5}
\end{figure}

Finally, when only one of the QDs is locally detuned, the minimal model at the  sweet spot in Eq. \eqref{Eq:delta_condition} predicts a quadratic splitting of the ground state degeneracy for $\delta n_g\neq0$, see Eq. \eqref{Eq:dE_splitting3}. For $\delta n_g=0$, instead, the sweet spot is achieved for equal CAR and COT amplitudes, such that Eq. \eqref{Eq:dE_splitting3} predicts that the degeneracy is preserved analogously to the non-interacting case. Surprisingly, the microscopic model gives the opposite result, namely a larger energy splitting for $\delta n_g=0$ than for $\delta n_g=0.1$, see Fig.~\ref{fig_4}(a,b). 

The difference between the results of the minimal and microscopic models can be understood by considering the MP for the left and right QDs. As shown in Figs.~\ref{fig_4}(c,d), detuning $\varepsilon_L$ away from the sweet spot mainly affects the MP at the right side. However, we note that also the MP at the left side is affected, mostly for $\delta n_g=0$ for the studied parameters. This effect is not captured by the minimal model and it is a consequence of the non-perturbative coupling between the QDs and the superconducting island, and the finite magnetic field considered. To illustrate this point, we show the ground state degeneracy lines in the $(\varepsilon_L,\varepsilon_R)$ map for two different values of the QD interactions, Fig.~\ref{fig_5}. For $U_\nu=0$, we find that the ground state splitting is weaker at the charge degeneracy point when detuning one of the QDs level, as predicted by the perturbative model~\eqref{Eq:Ham1}, Fig.~\ref{fig_5}(a). The tilt of the charge-degeneracy lines in the $(\varepsilon_L,\varepsilon_R)$ plane for $\delta n_g\neq0$ can be estimated from Eq.~\eqref{Eq:dE_splitting} in the perturbative regime and is a consequence of the different values of the COT and CAR amplitudes needed to reach the sweet spot that tend to shrink the different charge sectors. In contrast, Fig.~\ref{fig_5}(b) shows that the ground state splitting can be larger at the charge-degeneracy point than for a finite $\delta n_g$ for sizeable intradot interactions, $U_\nu$. In panel (b), this effect is evident when comparing the $\delta n_g=0.1$ degeneracy lines, which are almost parallel to the axes, to the degeneracy lines at the charge degeneracy points, which display more evident slopes. Therefore, intradot interactions $U_\nu$ and a finite Zeeman splitting may yield qualitative corrections on the position of the zero-bias peaks observed through transport experiments, that cannot immediately be explained through the perturbative model.

\section{Conclusions}

Minimal Kitaev chains provide a useful class of systems to probe the physics of Majorana modes in a controllable way. In this work, we analyzed the properties of QDs coupled via a floating superconductor that features a charging energy. Using a minimal model that describes the limit of weak coupling between the QDs and the superconducting island and a large Zeeman energy, we found the conditions for the onset of Majorana sweet spots, at and away from the island's charge-degeneracy points.

The existence of Majorana sweet spots is confirmed using a microscopic model that describes more realistically the degrees of freedom of the superconducting islands and the QDs. This microscopic approach confirms the stability of the sweet spots under perturbations of the QDs and provides an accurate description of the system beyond the perturbative regime.

Charging energy is an important element in many proposed devices and experiments based on superconducting nanowires engineered to host Majorana modes. 
The appearance of Majorana sweet spots in devices with floating superconducting islands opens for the possibility to engineer a future generation of interacting setups based on controllable PMMs. 
In particular, these devices could be used to integrate PMMs in transmon devices or to mimic
the physics of Majorana - Cooper pair boxes in controllable platforms. This suggests, in turn, alternative routes for the study of Majorana modes, with the possibility of reproducing some of the properties of topological qubits, such as as the Majorana tetron \cite{Plugge_NJP2017}, and to observe distinctive signatures of the Majorana non-locality as the topological Kondo effect \cite{Nitsch24}.

\section{Acknowlegements}
R.S.S acknowledges funding from the Horizon Europe Framework Program of the European Commission through the European Innovation Council Pathfinder Grant No. 101115315 (QuKiT), the Spanish Comunidad de Madrid (CM) ``Talento Program'' (Project No. 2022-T1/IND-24070), and the Spanish Ministry of Science, innovation, and Universities through Grants PID2022-140552NA-I00. V.B. acknowledges funding from the Romanian Ministry of Education and Research through Project No. 760122/31.07.2023 within PNRR-III-C9-2022-I9. M.B. has been supported by the Villum Foundation (Research Grant No. 25310). M.L. and M.N. acknowledge funding from the European Research Council (ERC) under the European Unions Horizon 2020 research and innovation programme under Grant Agreement No. 856526, the Swedish Research Council under Grant Agreement No. 2020-03412, and NanoLund.  L.M. has been supported by the research grant ``PARD 2023'', and ``Progetto di Eccellenza 23-27'' funded by the Department of Physics and Astronomy G. Galilei, University of Padua.

\appendix

\section{Perturbative regime: Crossed Andreev reflection and elastic cotunneling} \label{sec:pert}

In this appendix we derive the expression for the amplitudes of the elastic cotunneling $t_{\rm COT}$ and the crossed Andreev reflection $\Delta_{\rm CAR}$ by considering the perturbative limit ($t_\nu,t^{\rm SO}_\nu \ll \Delta - E_c$) of the microscopic model presented in Sec. \ref{Sec:microscopic}. For the sake of simplicity, we consider non-interacting QDs ($U_\nu=0$) and assume that the system is symmetric under the exchange of the left and right QDs, with $t_L=t_R\equiv t$ and $t_L^{\rm SO}=t_R^{\rm SO}\equiv t^{\rm SO}$.

As a first step, we consider the energies of the states of the superconducting island, as defined by the Hamiltonian ${H}_{\text{I}}$ in Eq. \eqref{H_SI}. We treat separately the charging energy, which depends on the number of Cooper pairs $N_c$, and the eigenenergy of the Bogoliubov quasiparticles. In particular, we disregard the Bogoliubov quasiparticle excitations with energies above the bulk gap of the superconductor and we consider exclusively the subgap states originating in the central region of the device as effect of the proximity-induced gap $\Delta$. Their eigenenergy is given by:
\begin{equation}
\tilde{\Delta} = \sqrt{\Delta^2 + \varepsilon_c^2}\,,
\end{equation}
such that the quadratic part of the Hamiltonian \eqref{H_SI} is diagonalized as $H_{\rm Bqp}=\sum_\sigma \tilde{\Delta} f^\dag_\sigma f_\sigma$, where we introduced the Bogoliubov modes:
\begin{align} \label{Bog1}
&f_{\Up } = u c_{\Up }\ee^{i\hat{\phi}/2} + v c^\dag_{\Dn }\ee^{-i\hat{\phi}/2}\,,\\
&f_{\Dn } = u c_{\Dn }\ee^{i\hat{\phi}/2} - v c^\dag_{\Up }\ee^{-i\hat{\phi}/2}\,,\label{Bog2}
\end{align}
with:
\begin{equation}
u = \sqrt{\frac{1}{2}+\frac{\varepsilon_c}{2\tilde{\Delta}}}\,,\quad v=\sqrt{\frac{1}{2}-\frac{\varepsilon_c}{2\tilde{\Delta}}}\,.
\end{equation}

When $E_c < \tilde{\Delta}$, the low energy states of the SC island are all characterized by an even number $2N_c$ of electrons,  independently of the value of $n_g$. This reflects the fact that the energy cost of occupying any Bogoliubov mode is larger than the difference in charging energy between even and odd states, even when $n_g$ is an odd integer $n_{g,0}$.

The condition $E_c < \tilde{\Delta}$ is indeed necessary for the charging energy to assume the form \eqref{H_I1} and it is in this strong pairing regime that both COT processes and CARs are properly defined when we consider the weak tunneling regime $t,t^{\rm SO} \ll \tilde{\Delta} - E_c$. 

COT processes amounts to the coherent tunneling of an electron from one external QD to the other, mediated by a Bogoliubov state. In this process, the number of Cooper pairs on the superconducting island is left unchanged.

The crossed Andreev reflection, instead, corresponds to a process in which a pair of electrons, each coming from a different QD, enters the superconducting island by increasing the number of Cooper pairs by one. Also in this case, the process is mediated by the Bogoliubov state. 
In the following, we consider for simplicity a charge offset $0\le n_g \le 2$, such that the lowest lying states of the superconductor display $N_c=0$ or $1$. All the results are invariant by shifting $n_g \to n_g\pm 2$.

We derive first the cotunneling amplitude. By inverting Eqs. \eqref{Bog1} and \eqref{Bog2}, we get:
\begin{equation}
c^\dag_{\sigma} = \left( u f^\dag_{\sigma } -s_\sigma v f_{\bar{\sigma} }\right)\ee^{i\hat{\phi}/2} \,.
\end{equation}
Expressed in terms of the Bogoliubov modes $f_\sigma$ the tunneling Hamiltonian \eqref{ham_hop} reads:
\begin{multline}
H_T = 
  \sum_{\nu,\sigma} t\left[\left(u f^\dag_{\sigma} - s_\sigma v f_{\bar{\sigma}}\right)
 d_{\nu \sigma}\ee^{i \hat{\phi}/2} + {\rm H.c.}\right]+\\
 \sum_{\nu,\sigma}t^{\rm SO} s_\nu \left[\left(u s_\sigma f^\dag_{\bar{\sigma}} +  v f_{\sigma}\right)
 d_{\nu \sigma}\ee^{i \hat{\phi}/2} + {\rm H.c.}\right].
\end{multline}
In these expressions, the operator $\ee^{i\hat{\phi}/2}$ increases the charge of the SC island by one. 

In the following, we assume that the Zeeman energy of the QDs is strong ($E_Z \gg t,t^{\rm SO}$) and polarizes them, in order to retrieve the description in Eq. \eqref{HQDs}. In particular, we consider a scenario in which the occupation of the spin $\Dn$ states is suppressed. We also set $\varepsilon_L=\varepsilon_R=\varepsilon$ for simplicity. Under these assumptions, the elastic cotunneling at second order is described by:
\begin{multline} \label{cot1}
H_{\rm COT} = \sum_{ \nu^{\prime}, \nu^{\prime\prime}} -d^\dag_{\nu^{\prime \prime} \Up}  \frac{\left[\left(t^{\rm SO}\right)^2  s_{\nu^{\prime \prime}}s_{\nu^{\prime}} +t^2 \right] u^2}{\tilde{\Delta} + E_{c}\left(1+2N_{\rm{I},0}-2n_g \right) -\varepsilon} d_{\nu^{\prime } \Up } \\
-  d_{ \nu^{\prime \prime} \Up } \frac{\left[\left(t^{\rm SO}\right)^2  s_{\nu^{\prime \prime}}s_{\nu^{\prime}} +t^2 \right] v^2}{\tilde{\Delta} +   E_{c}\left(1-2N_{\rm{I},0}+2n_g \right)+\varepsilon}d^\dag_{ \nu^{\prime } \Up }\,,
\end{multline}
where we labelled by $N_{\rm{I},0}$ the value of the total charge of the SC island $\hat{N}_{\rm I}$ in the initial state. In the limit $E_c=0$ we recover the results in Ref.~\cite{Liu_PRL2022}. 
For $\nu^{\prime} \neq \nu^{\prime \prime}$, this expression corresponds to the cotunneling term in Eq. \eqref{Heff} and:
\begin{multline}
t_{\rm COT}(N_{\rm{I}})= \frac{\left(t^{\rm SO}\right)^2 - t^2}{(\tilde{\Delta} + E_c )^2 - \left[E_c\left(2N_{\rm{I}}-2n_g\right)-\varepsilon \right]^2} \cdot\\
\left[2\left(n_g-N_{\rm{I}} \right)E_c + \varepsilon +\varepsilon_c \left(1+\frac{E_c}{\tilde{\Delta}} \right) \right].
\end{multline}
For $\nu^{\prime}=\nu^{\prime \prime}$, instead, we obtain a shift of the QD potential that corresponds to
\begin{multline} \label{deltaepsilon}
\delta \varepsilon(N_{\rm{I}})= -\frac{\left(t^{\rm SO}\right)^2 + t^2}{(\tilde{\Delta} + E_c )^2 - \left[E_c\left(2N_{\rm{I}}-2n_g\right)-\varepsilon \right]^2} \cdot\\
\left[2\left(n_g-N_{\rm{I}} \right)E_c + \varepsilon +\varepsilon_c \left(1+\frac{E_c}{\tilde{\Delta}} \right) \right],
\end{multline}
such that we can define the renormalized energy levels of the QDs as:
\begin{equation} \label{epsilon}
\epsilon_\nu(N_{{\rm I}}) = \varepsilon_\nu +\delta\varepsilon(N_{{\rm I}})\,.
\end{equation}
When considering $n_g \sim 1$, thus charge values $N_{{\rm I}}=0,2$, it is easy to check that the shift of the QD energy caused by the different occupation of the SC island is approximately of order $4\left[ \left(t^{\rm SO}\right)^2 + t^2 \right]E_c/ \tilde{\Delta}^2$.

Analogous calculations can be performed for the CAR amplitude. 
CAR processes connect charge configurations with different energies unless the relation $\epsilon(N_{\rm I}=0) = -\delta E_c/2$ holds, which generalizes Eq. \eqref{constr1} by including the explicit dependence on $N_{\rm I}$. A perturbative approach to determine the CAR amplitude when the degeneracy between the charge configurations with $N_{\rm I}=0$ and $N_{\rm I}=2$ is lifted can be applied, in general, if $\tilde{\Delta} \gg \sqrt{\left(t^{\rm SO}\right)^2 + t^2} \gtrsim E_c$. Under this condition we obtain that CAR defines an effective p-wave pairing between the QDs of the form

\begin{equation}\label{carH}
H_{\rm CAR} = -\frac{2t^{\rm SO}t\sqrt{\tilde{\Delta}^2-\varepsilon_c^2}}{\tilde{\Delta}\left(\tilde{\Delta}-E_c\right)}  \ee^{i\hat{\phi}} d_{R \Up} d_{L \Up}  + {\rm H.c.}\,,
\end{equation}
which provides the expression for the amplitude $\Delta_{\rm CAR}$. We observe that, differently from the cotunneling amplitude and the energies $\epsilon_\nu$, $\Delta_{\rm CAR}$ does not depend on the charge $N_{\rm I}$.

Based on the previous results, we can refine the general arguments of Sec. \ref{sec:minimalModel} by considering the role of the charge of the SC island in more detail. As seen in Sec. \ref{Sec:microscopic}, depending on the microscopic parameters, the ground state of the system may display different particle numbers. In particular, when focusing on $0<n_g<2$ and $|\varepsilon_\nu| \ll \Delta$, the most typical scenario is to have even ground states with a total electron number $N=2$ and odd ground states with $N=1,3$.
To search for zero-energy modes, therefore, we must separately consider either the degeneracies between states with $N=2$ and $N=1$, or between $N=2$ and $N=3$.

In the even sector ($N=2$) the low-energy Hamiltonian \eqref{He} presents effective QD energies $\epsilon_\nu(N_{\rm I}=0)$ related to the state $\ket{11,N_c=0}$, as defined in Eq. \eqref{epsilon}. Therefore, to obtain effective eigenstates of the kind \eqref{estate}, we must impose the condition:
\begin{equation} \label{constr}
\epsilon_L(0) = \epsilon_R(0) = -\frac{\delta E_c}{2} = 2E_c\left(1-n_g\right)\,,
\end{equation}
which corresponds to Eq. \eqref{constr1} for $N=2$.
This condition guarantees that the charge states connected by CAR processes have the same energy.

In the odd sectors, the effective eigenstates always have the form in Eq.~\eqref{ostate}. However, the $N=1$ sector has energies determined by $\epsilon_\nu(N_{\rm I}=0)$, whereas the $N=3$ sector displays effective QD energies $\epsilon_\nu(N_{\rm I}=2)$.

\begin{figure}[t!]
\includegraphics[width=\columnwidth]{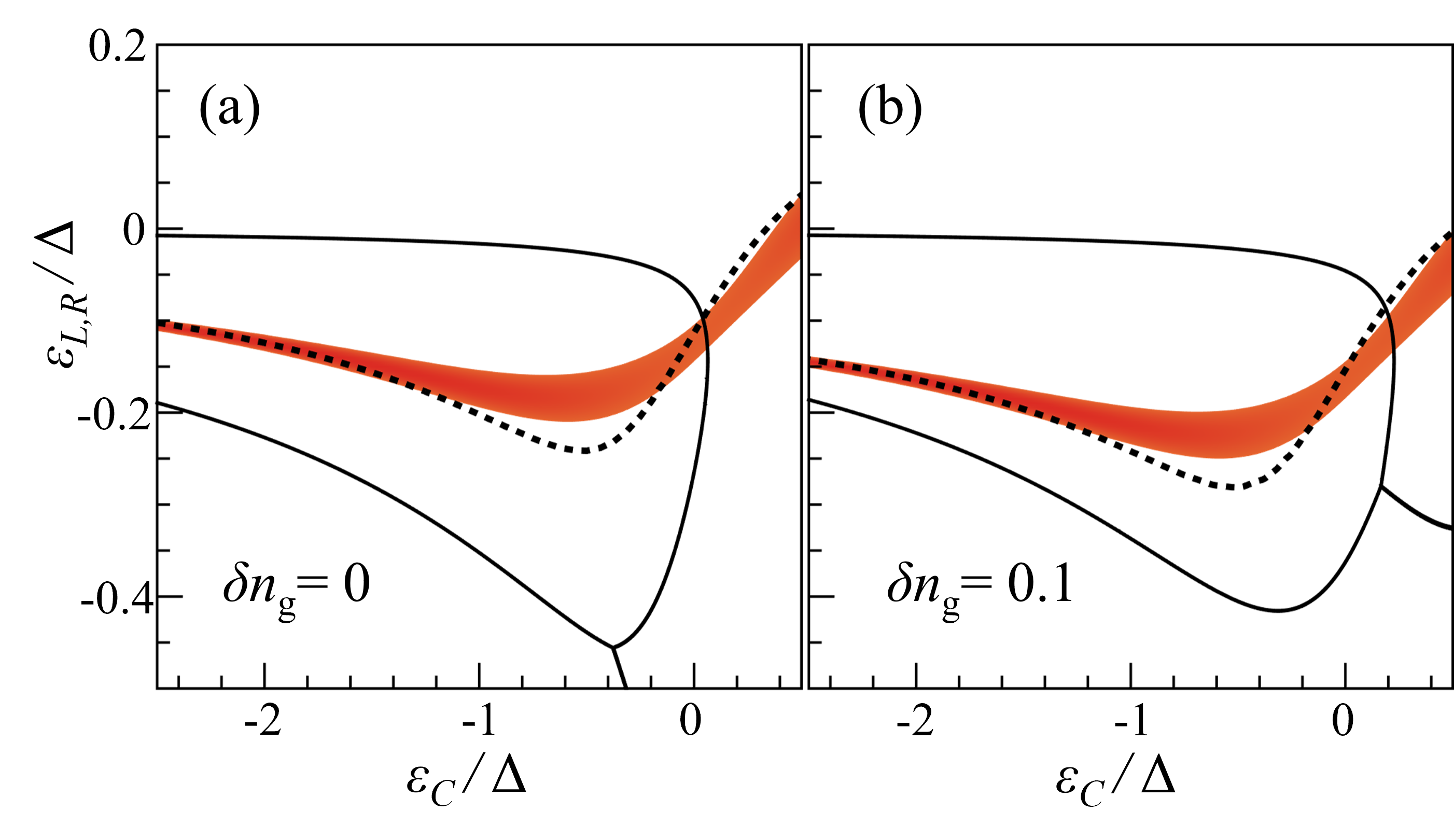}
    \caption{Same as Fig.~\ref{fig_2}(a,c) but for $U=0$ and displaying only the region $|\text{MP}_{21}|>0.9$ (in red). The dashed lines correspond to the constraint \eqref{constr}, with $N_c=1$ for $n_{g,0}=1$. Panel (a) refers to the charge degeneracy point $n_g=n_{g,0}$, whereas panel (b) displays data for $n_g=n_{g,0}+0.1$. The discontinuity in the red region in panel (b) originates from a crossing of the excited eigenstates in the sector with $n_{g,0}+2$ electrons. We observe that the peak of the MP is well captured by the perturbative relations \eqref{epsilon} and \eqref{constr}.
    }
\label{Ap:fig_1}
\end{figure}

Therefore, when the condition \eqref{constr} is fulfilled, the degeneracy of the even ($N=2$) and odd ($N=3$) states is achieved for
\begin{equation} \label{constr2}
   \left|\Delta_{\rm CAR}\right|=\left|t_{\rm COT}(N_{\rm I}=2)\right|-\epsilon(N_{\rm I}=2)\,.
\end{equation}
The difference between this relation and Eq. \eqref{Eq:delta_condition} lies in the difference between $\epsilon(0)=-\delta E_c/2$ [from the constraint \eqref{constr}] and $\epsilon(2)$, which, in turn, is determined by the different shifts in Eq. \eqref{deltaepsilon}. 
Eq. \eqref{Eq:delta_condition} neglects this difference. 
However, the peak of the Majorana polarization that we analyzed in the main text depends only on the ground state wavefunctions and not on their energy degeneracy. Therefore, it is approximately determined based on the relation \eqref{constr} only. Indeed, the peak of the polarization $\text{MP}_ {21}$, which reflects how closely the wavefunction \eqref{estate} approximates the ground state, precisely follows the condition \eqref{constr} for low values of the tunneling rates (see the dashed lines in Fig. \ref{Ap:fig_1}).

Finally, we observe that the different values of $\epsilon$ for states with different charges yield a weak dependence of the energy differences in Eqs. (\ref{Eq:dE_splitting}-\ref{Eq:dE_splitting3}) on the occupation of the central island.

\section{{Equivalent BCS microscopic model}}
\label{app_BCS}

As the setup under investigation in this work involves a single superconducting island, it is possible to gauge away the superconducting phase degree of freedom by using the conservation of the total particle number
\begin{equation}
    \hat{N}=\hat{N}_{\rm I}+\sum_{\nu=L,R} \sum_{\sigma=\uparrow,\downarrow} d_{\nu\sigma}^\dagger d_{\nu\sigma}~.
\end{equation}
To this purpose we apply the following gauge transformation,
\begin{equation}
\label{gauge_transf}
    c^\dagger_\sigma \rightarrow c^\dagger_\sigma \ee^{i\hat{\phi}/2}~,~d^\dagger_{\nu\sigma} \rightarrow d^\dagger_{\nu\sigma} \ee^{i\hat{\phi}/2}~,
\end{equation} 
which leaves the tunneling term of the Hamiltonian in Eq. \eqref{ham_hop} invariant, but allows us to
rewrite the Hamiltonian of the superconducting island \eqref{H_SI} in the BCS picture:
\begin{equation}\label{H_SI_bcs}
\begin{aligned}
    {H}_{\text{I,BCS}}=& \sum_{\sigma=\uparrow \downarrow} \varepsilon_c c_{\sigma}^{\dagger} c_{\sigma}- (\Delta c_{\uparrow}^{\dagger} c_{\downarrow}^{\dagger}+\text{H.c.})\\
    &+E_c \left(\hat{N}-\sum_{\nu,\sigma} d_{\nu\sigma}^\dagger d_{\nu\sigma}-n_g\right)^2,
    \end{aligned}
\end{equation}
where $\hat{N}$ is a conserved quantity, such that the total Hamiltonian of the system explicitly acquires a block diagonal form.

In the original picture, i.e. before performing the transformation in Eq.~(\ref{gauge_transf}), a generic state with an even particle number $N=2n$ may be expanded schematically as
\begin{equation}
\begin{aligned}
    |2n\rangle =& ~|N_c=n\rangle+ c^\dagger d^\dagger  |N_c=n-1\rangle\\
    &+ c^\dagger c^\dagger d^\dagger d^\dagger  |N_c=n-2\rangle+\dots~,
\end{aligned}
\end{equation}
where $|N_c\rangle$ denotes the state with $N_c$ Cooper pairs in the SC condensate and no electrons on the QDs. In the BCS picture, this becomes a superposition of various even numbers of gauge-transformed fermion operators,
\begin{equation}
\begin{aligned}
    |2n\rangle =&\, (1+ c^\dagger d^\dagger + c^\dagger c^\dagger d^\dagger d^\dagger+\dots)~|N_c=n\rangle\\
    \equiv &\,|E_{2n}\rangle~,
\end{aligned}
\end{equation}
with $|N_c=n\rangle$ acting as their vacuum state. In the last equality, we used the explicit notation $|E\rangle$ to emphasize the even parity of the BCS combination. Similar considerations apply for the case of an odd particle number state equivalent to an odd-parity BCS combination, i.e. $|2n+1\rangle\equiv|O_{2n+1}\rangle$.

The combinations that appear in the MP of Eq.~(\ref{MP}), written in the original picture as, e.g.
\begin{equation}
    w_{\nu\sigma}=\langle 2n|\left(\ee^{i\hat{\phi}}d_{\nu\sigma}+d_{\nu\sigma}^\dagger\right)|2n-1\rangle~,
\end{equation}
is simplified in the BCS picture as
\begin{equation}
    w_{\nu\sigma}=\langle E_{2n}|\left(d_{\nu\sigma}+d_{\nu\sigma}^\dagger\right)|O_{2n-1}\rangle~,
\end{equation}
when expressed in terms of gauge-transformed fermion operators. Here, $|E_{2n}\rangle$ and $|O_{2n-1}\rangle$ are the ground states of $H_\text{BCS}(2n)$ and $H_\text{BCS}(2n-1)$ in their even and odd fermion-partity subspaces.  

We finally observe that while the information on the particle-number is fully contained in all the above expressions, the BCS picture is preferred for the numerics due to its computational simplicity.

\section{Charging energy effects}\label{App::c}
\begin{figure}[ht!]
\includegraphics[width=\columnwidth]{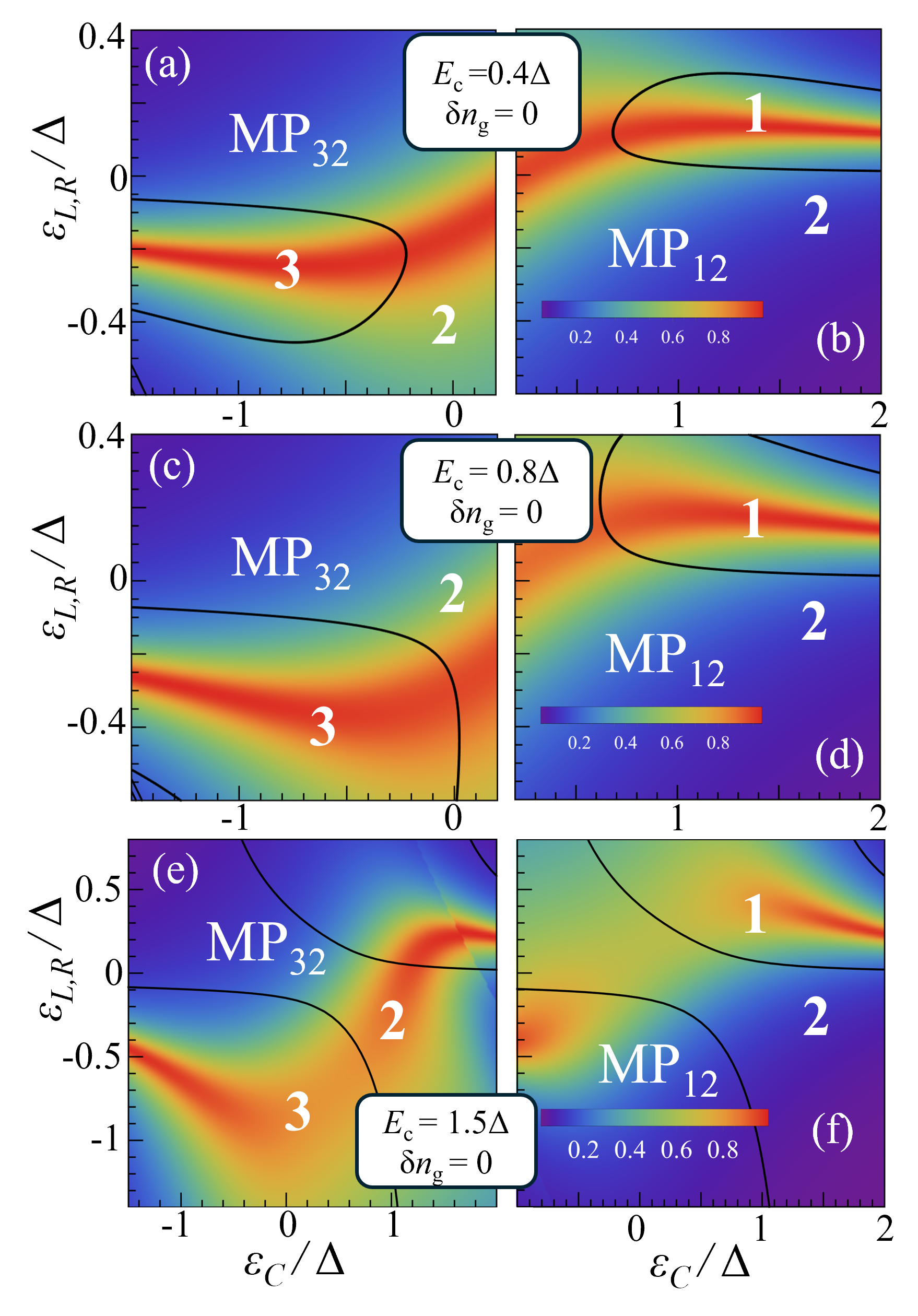}
    \caption{Same as Fig.~\ref{fig_2}(a,b) but for larger $E_c$. Regions with ground state particle numbers $k=1,2,3$ are indicated by white labels.}
\label{Ap:fig_1b}
\end{figure}
In this appendix, we analyze the effect of the charging energy of the island, $E_c$, focusing on the existence of Majorana sweet spots. In Fig.~\ref{Ap:fig_1b}, we show the calculated MP and the ground state degeneracy conditions (black lines) for increasing $E_c$ values, using the full microscopic model presented in Sec.~\ref{Sec:microscopic}. We first focus on the island's charge-degeneracy points, where the effects of the charging energy are predicted to be the smallest. These results complement the ones presented in Fig.~\ref{fig_2} of the main text. As shown for the chosen parameters, the system features Majorana sweet spots for all studied $E_c$, characterized by a degenerate ground state and high MP values. However, the value of MP at the sweet spot decreases monotonically with increasing $E_c$. This decrease of MP is correlated to the enhancement of the occupations $\left\langle f^\dag_\sigma f_\sigma \right \rangle$, with $f$ being the operator associated with  quasiparticle excitations in the SC island, at the charge degeneracy point, which spoils the interpretation of the system as a minimal Kitaev chain. The introduction of Bogoliubov quasiparticles in the island implies indeed that the occupation of the QDs is no longer directly related to the fermionic parity of the system. This effect is particularly significant at the charge degeneracy point for systems with an odd number of total electrons when $E_c>\Delta$: for the data displayed in Fig. \ref{Ap:fig_1b}(c) the probability of occupying Bogoliubov states for $k=+2$ approaches $0.7$ in proximity of the maximum of MP along the degeneracy line.

\begin{figure}[ht!]
\includegraphics[width=0.8\columnwidth]{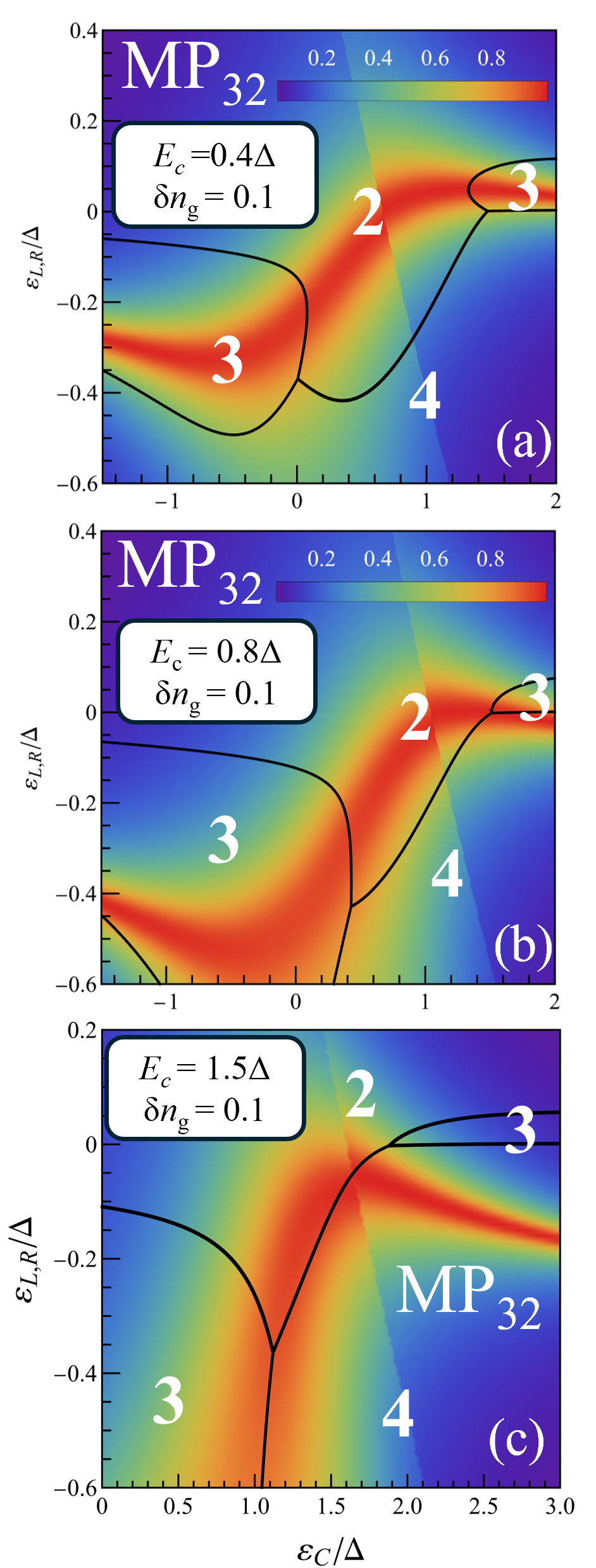}
    \caption{Same as Fig.~\ref{fig_2}(c) but for larger $E_c$. Regions with ground state particle numbers $k=2,3,4$ are indicated by white labels.}
\label{Ap:fig_2}
\end{figure}

In Fig.~\ref{Ap:fig_2} we display the corresponding results away from the charge-degeneracy point, $\delta n_g\neq0$. In this situation, the MP acquires, in general, peak values that are larger than the charge-degeneracy point for the same $E_c$. This is consistent with a lower occupation $\left\langle f^\dag_\sigma f_\sigma\right\rangle$ of the Bogoliubov excitations. Moreover, we note the appearance of additional charge sectors, labelled using white numbers. These additional charge sectors lead to triple degeneracies: points where ground states with three different values of the total number of particles are degenerate. These triple degeneracies can appear in regions with a high MP between two of the states. Their characterization is an interesting open question which is suggestive of the onset of weak parafermionic modes \cite{Bozkurt_arXiv2024}, although it seems unlikely that these zero-energy modes are localized for a generic choice of the parameters $E_c$ and $n_g$.

Finally we observe that the discontinuities displayed by ${\rm MP}_{21}$ in Fig. \ref{Ap:fig_2} are caused by a crossing of the excited states of the odd sectors.

\section{Multiple subgap states in the island}
\label{app_surrogate}
We consider in this section a microscopic approach to the SI beyond the single Andreev bound state picture of the main text, and based instead on the multi-level surrogate model \cite{Baran_PRB2023, Baran2024Jun}. In this context, the full quasi-continuum of levels of the superconducting island (SI) is replaced by a small number $\tilde{L}$ of BCS surrogate orbitals that optimally reproduce the SI-QD hybridization function. Such a description may actually be necessary for correctly addressing certain physical aspects of Andreev molecular states, such as the subtle competition between the spin-singlet and triplet states in QD-SI-QD devices \cite{Baran2024Jun, Baran2024Aug, Bacsi2023Sep, Zalom2024Jul}.  For such multi-level models, the increase in computational complexity may be well mitigated by employing the reduced basis method \cite{Baran2023Apr, Herbst2022Apr, Brehmer2023Apr}.

Concretely, we consider a minimal $\tilde{L}=2$ model of the SI for which Eq.~(\ref{H_SI}) is modified as
\begin{equation}\label{H_SI_L=2}
\begin{aligned}
    {H}_{\text{I}} =& \sum_{j=1}^2\sum_{\sigma=\uparrow \downarrow} \varepsilon_{j\sigma} c^{\dagger}_{j\sigma}  c_{j\sigma}- \sum_{j=1}^2(\Delta c_{j\uparrow}^{\dagger} c_{j\downarrow}^{\dagger} \ee^{-i{\hat{\phi}}}+\text{H.c.})\\
    &+E_c \left(\hat{N}_{\rm I}-n_g\right)^2~.
    \end{aligned}
\end{equation}
Here, the single-particle energies are given by $\varepsilon_{1,2;\sigma}=\pm\xi+\mu_{\text{SC}}$, where the level distance $\xi=1.31\Delta$ has been obtained by the hybridization function fit at vanishing chemical potential $\mu_{\text{SC}}$ in the proximitized region \cite{Baran_PRB2023}.

Furthermore, we consider the best-case scenario where both levels are coupled symmetrically to each QD. A more realistic description would need to account for the microscopic randomness of the system by the use of arbitrary coupling parameters \cite{Malinowski2023Mar,Baran2024Jun,Kurtossy2024Jun}. The latter's distribution would thus define a measure for the overlap between the Yu-Shiba-Rusinov states of the left/right QDs. The main signatures of the Andreev molecular physics depend smoothly on this overlap, being absent when the QDs are effectively decoupled (vanishing overlap) and being most visible in the present symmetric coupling situation (unit overlap). 

\begin{figure}[t!]
\includegraphics[width=\columnwidth]{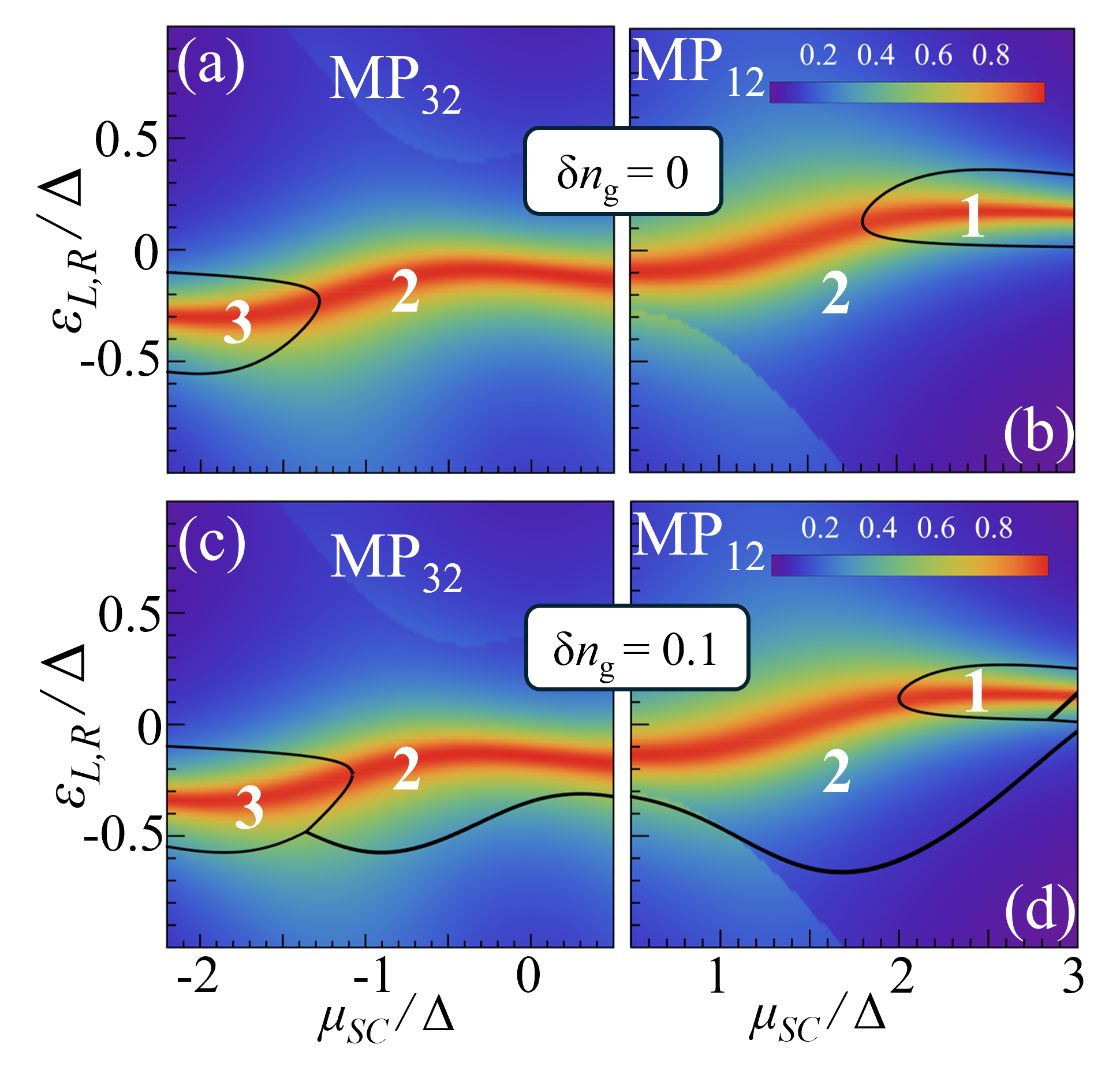}
    \caption{Same as Fig.~\ref{fig_2} but for the $\tilde{L}=2$ surrogate model. Regions with ground state particle numbers $k=1,2,3$ are indicated by white labels.  }
\label{Ap:fig_3}
\end{figure}

The results of the $\tilde{L}=2$ model are shown in Fig.~\ref{Ap:fig_3} below for the tunneling amplitudes between each SI level and each QD taken as $t=0.5\Delta$, corresponding to a QD-SI tunneling rate $\Gamma=0.2\Delta$ (with the other parameters kept the same as in the single-level case). While the presence of an extra Andreev level in the SI is seen to induce some slight changes in the stability diagram, the main qualitative PMM signatures remain unaffected. The single-level model considered in the main text is thus confirmed to account well for the relevant PMM physics.

\bibliography{bibliography.bib}

\end{document}